\documentclass[twocolumn,secnumarabic,amssymb,nobibnotes,aps,pre,showpacs,superscriptaddress]{revtex4-1}
\usepackage[english]{babel}
\usepackage{graphicx}
\usepackage{amsmath}
\usepackage{amsfonts}
\usepackage[usenames,dvipsnames]{color}
\usepackage{hyperref}
\usepackage{blindtext}
\hypersetup{colorlinks}
\hypersetup{linkcolor=black, citecolor=blue, urlcolor=blue}
\usepackage[utf8]{inputenc}
\usepackage{color}
\setcitestyle{super}

\begin{document}
\title{Screened Coulomb interactions of general macroions with nonzero particle volume}
\author{Jeffrey C. Everts}
\email{jeffrey.everts@gmail.com}

\address{Faculty of Mathematics and Physics, University of Ljubljana, Jadranska 19, 1000 Ljubljana, Slovenia}

\date{\today}

\begin{abstract}
A semianalytical approach is developed to calculate the effective pair potential of rigid arbitrarily shaped macroions with a nonvanishing particle volume, valid within linear screening theory and the mean-field approximation. The essential ingredient for this framework is a mapping of the particle to a singular charge distribution with adjustable effective charge and shape parameters determined by the particle surface electrostatic potential. For charged spheres this method reproduces the well-known Derjaguin-Landau-Verwey-Overbeek (DLVO) potential. Further exemplary benchmarks of the method for more complicated cases, like tori, triaxial ellipsoids, and additive \mbox{torus-sphere} mixtures, leads to accurate closed-form integral expressions for \emph{all} particle separations and orientations. The findings are relevant for determining the phase behaviour of macroions with experiments and simulations for various particle shapes. 
\end{abstract}

\maketitle

%Effective pair potentials are of prime importance in describing the phase behaviour of many-body systems. Not only are a lot of systems described by pair interactions, while for only a few three body is important, but also by integrating out certain degrees of freedom that one are not directly interested, the complexity of the system can be significantly reduced. Effective interactions that were very succesfull and lead to unexpected sciientific experiemntsn describing the physics are for example the phonon-mediated attractive electron-electron interaction in the BCS theory for superconductors, BEC, etc while the bair interaction is repulsive, the effective interaction is attractive. In soft matter integrating out polymer degrees of freedom yielded for example the depletion potential, solvent-mediated potentials, these effective pair potentials are also very useful in describing many-body physics by the use of for example simulations, while in quantum mechanics simulations are limited because of the fermion sign problem.
\section{Introduction}
Screened Coulomb interactions of electronic or ionic nature are ubiquitous in quantum-mechanical and classical systems, such as strongly correlated electron matter \cite{Werner:2010, Mauri:2019}, chemical bonds \cite{LJ:1951, Wang:2014}, superconductors \cite{Parendo:2005}, proteins \cite{Schreiber:1996}, liquid crystals \cite{Musevic:2002, Mundoor:2016, Everts:2020}, DNA \cite{Podgornik:1997, Kornyshev:2007, Lee:2011}, graphene \cite{dasSarma:2007}, lipid membranes \cite{Goldstein2:1990, Andelman:1995} , supercapacitors \cite{Miller:2008, Kondrat:2010}, microfluidics \cite{Bazant:2004}, and dusty plasmas \cite{Thomas:1994, Hopkins:2005}. A general understanding of electrostatic screening in various geometrical settings is needed, considering that many of these systems have a complex geometry. In particular, classical charge-screened particles of various shapes and surface functionalities can be experimentally synthesised and characterised today in great detail \cite{Glotzer:2007, Boles:2016}, however, theoretical understanding of effective particle interactions is lagging behind as it is difficult to account for finite particle volume and non-spherical particle shape. This paper is aimed at bridging the gap between the available experimental and theoretical toolkits.

In order to express the system in solely the degrees of freedom of interest --- such as the positions and orientations of specific particles ---  it is useful to integrate out the ``fast" charge degrees of freedom, which leads to an effective description in terms of electrostatic screening. In free-electron like metals this procedure leads to Friedel oscillations, electron density modulations near a solid-fluid interface and around impurities \cite{Kohn:1970,Einstein:1973,Friedel:1958}, which is a canonical example of an emergent phenomenon caused by screening. In classical systems \cite{Debye:1923} and sufficiently dilute quantum systems \cite{Thomas:1927, Fermi:1928}, screening is often associated with the damped spatial decay of the electrostatic potential (and thus the effective pair potential), that for point particles and for low voltages compared to the thermal energy, has the Yukawa form, $\sim\exp(-\kappa r)/r$, with $\kappa^{-1}$ being the Debye screening length and $r$ being the radial distance, as opposed to the bare Coulomb case $\sim 1/r$. Of special interest are spherical charged particles, with radius $a$  in the colloidal (sub)micron regime, dispersed in ion-containing liquids, because of their tunable charge and screening properties \cite{Blaaderen:2003}. Integrating out the degrees of freedom of the smaller ions results in a Yukawa-type effective sphere-sphere potential with a prefactor that depends on the particle charge and, unlike for point particles, also the salt concentration via $\kappa a$ arising from the ion-impenetrable particle hard core \cite{Derjaguin:1948, VerweyOverbeek}. This so-called Derjaguin-Landau-Verwey-Overbeek (DLVO) potential \footnote{I will only consider the electrostatic part of the full DLVO potential in this paper.} is an essential theoretical tool for understanding the behaviour of charge-stabilised colloids \cite{Belloni:2000}, even for out-of-equilibrium suspensions \cite{Zaccone:2009}.

For nonspherical shapes, the screened-electrostatic pair interaction is only analytically known in a few cases  for all particle configurations, even within linear screening theory. However, some studies exist for disks \cite{Trizac:2000, Trizac:2002, Agra:2004, Jabbari:2014}, rods \cite{Brenner:1974, Eggen:2009}, spheroids \cite{Hsu:1997, Tellez:2010, Schiller:2011, Nagele:2012, Cruz:2015}, or helices \cite{Kornyshev:1997}, where the potential is sometimes calculated only for infinitely long, thin, or ion-penetrable particles, restricted particle configurations or orientation-averaged interactions \cite{Hsu:1998}. The difficulty in finding analytical solutions lies in the finite ion-impenetrable particle volume which complicates matching the series expansion solution (if it is even available for the geometry under consideration) of the unscreened potential inside and the screened potential outside the particle via the boundary conditions. However, when the pair potential would be known, one does not need to numerically solve the three-dimensional Poisson(-Boltzmann) equation for every single particle configuration at each simulation step, as in Refs. \cite{Fushiki:1992, Hansen:1992, Hallez:2014} for simulations of charged colloids. Instead, computationally less expensive simulations with effective pair potentials can be used, and by mapping to cell models, even charge regulation can be incorporated \cite{Everts:2016b}. The lack of availability of accurate pair potentials might explain why fewer phase behaviour studies are known for complex-shaped charged particles \cite{Dijkstra:2003, Jabbari:2008, Leunissen:2015} than for charge-neutral hard particles \cite{Graaf:2011, Gantapara:2013, Glotzer:2011, Glotzer:2012, Nijs:2015, Dussi:2016}.

In this paper, I devise a framework to semianalytically approximate effective interactions between (not necessarily equal) charged finite-size particles with not necessarily spherical shape. By mapping particles to singular charge distributions (i.e. expressed by $\delta$ functions), I find a straightforward and accurate evaluation of the interaction free energy for \emph{arbitrary} interparticle separations and orientations. For spheres,  this method reproduces DLVO theory, and hence a similar level of approximation is expected such as weak double layer overlap \cite{Carnie:1993}. After discussing spheres,  I apply the framework to more complicated shapes, such as tori and ellipsoids.

\begin{figure}[t]
\centering
\includegraphics[width=0.44\textwidth]{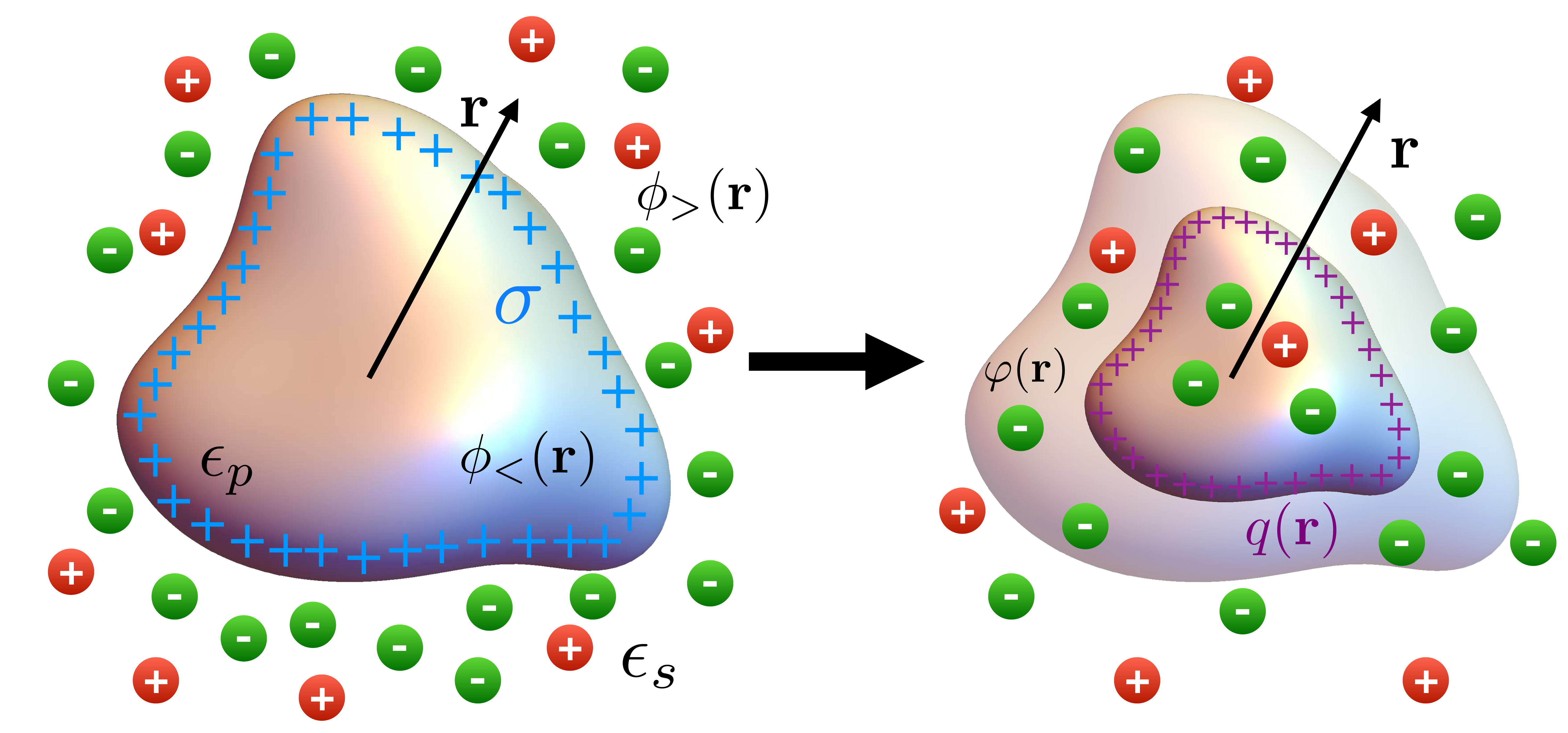} 
\caption{Scheme of mapping an arbitrary ion-impenetrable particle with surface charge density $\sigma$ to an effective ion-penetrable charge distribution $q({\bf r})$, that can be a point, line or surface charge.}
\label{fig:scheme}
\end{figure}

\section{General theoretical framework}
To set up the theoretical framework, I consider an ion-impenetrable charged particle of arbitrary shape with dielectric constant $\epsilon_p$, surface $\mathcal{P}$, and interior volume $\mathrm{int}(\mathcal{P})$, immersed in a structureless solvent with dielectric constant $\epsilon_s$ and Bjerrum length $\ell_B=\beta e^2/(4\pi\epsilon_0\epsilon_s)$, with $e$ being the proton charge, $\epsilon_0$ being the vacuum permittivity, and $\beta^{-1}=k_BT$, with $k_B$ being the Boltzmann constant and $T$ being temperature; see the scheme in Fig. \ref{fig:scheme}. For simplicity, I focus on a $1:1$ salt, such that the mean-field approximation is valid. By thermally averaging over the ions, an inhomogeneous electrostatic potential $\phi({\bf r})/(\beta e)$ describes the electric double layer that is formed around the particle. I split the total electrostatic potential as a contribution inside the particle $\phi_<({\bf r})=\phi({\bf r})|_{{\bf r}\in\mathrm{int}(\mathcal{P})}$, and a contribution outside the particle  $\phi_>({\bf r})=\phi({\bf r})|_{{\bf r}\notin\mathrm{int}(\mathcal{P})}$. For linear screening, there is the condition on the dimensionless electrostatic potential $|\phi({\bf r})|\ll 1$, meaning that I only consider potentials much smaller than the thermal voltage.  Within the mean-field approximation, $\phi_<({\bf r})$ and $\phi_>({\bf r})$ are given by the Laplace and Debye-H\"{u}ckel (DH) equations, respectively,
\begin{gather}
\nabla^2\phi_\mathrm{<}({\bf r})=0, \quad
\nabla^2\phi_\mathrm{>}({\bf r})=
\kappa^2\phi_\mathrm{>}({\bf r}),
\label{eq:DH}
\end{gather}
with $\kappa^{-1}=(8\pi\ell_B\rho_s)^{-1/2}$ being the Debye screening length where $\rho_s$ is the reservoir salt concentration, and I enforce continuity $\phi_\mathrm{<}({\bf r})=\phi_\mathrm{>}({\bf r})$ for ${\bf r}\in \mathcal{P}$, and constant-charge boundary condition on the particle surface with outward normal $\hat{\boldsymbol{\nu}}$ and surface charge density $e\sigma$,
\begin{equation}
{\hat{\boldsymbol{\nu}}}\cdot\Big[\epsilon_p\nabla\phi_\mathrm{<}({\bf r})-\epsilon_s\nabla\phi_\mathrm{>}({\bf r})\Big]/\epsilon_s=4\pi \ell_B\sigma, \quad {\bf r}\in\mathcal{P}.
\label{eq:BC}
\end{equation}
The finite particle volume complicates matching ${\phi}_>({\bf r})$ and ${\phi}_<({\bf r})$ through the boundary conditions \cite{Tellez:2010}, although formal exact multiple-scattering expansions exist \cite{Goldstein:1990}. Sometimes Eq. \eqref{eq:DH} is not even separable in certain coordinate systems, which further complicates finding analytical solutions. For example, due to the non-separability of the Helmholtz equation (DH with imaginary $\kappa$) in toroidal coordinates, the complete solution can only be expressed in terms of lengthy toroidal wave functions  \cite{Weston:1958, *Weston:1960}. However, some approximations for the double layer around a torus exist \cite{Blum:2005, Andreev:2006, *Andreev:2007}.

The first central message of this paper is that an ion-impenetrable charged particle can be mapped to a singular charge distribution $q({\bf r})$ \emph{without} a particle hard core, described by the DH equation $(\nabla^2-\kappa^2){\varphi}({\bf r})=-4\pi\ell_Bq({\bf r})$, such that the outside potential is approximated to a very high accuracy by $\phi_\mathrm{>}({\bf r})\approx\varphi({\bf r})$, schematically shown in Fig. \ref{fig:scheme}. Consequently, $\varphi({\bf r})$ can be expressed as the convolution of $q({\bf r})$ with the DH Green's function,
\begin{equation}
\varphi({\bf r})=\ell_B\int d{\bf r}'q({\bf r}')\frac{\exp\left(-\kappa|{\bf r}-{\bf r}'|\right)}{|{\bf r}-{\bf r}'|}.
\label{eq:DHconv}
\end{equation}
Here, $q({\bf r})$ can either be a point, line, or surface charge distribution and has to be parametrised with the same symmetry of the particle. The salt-dependent shape and magnitude of $q({\bf r})$ can be determined by matching, e.g. the numerically or analytically obtained surface potential. The benefit of this method is that the \emph{same} $q({\bf r})$ for a single particle, enters the approximate analytical expression for the charge-screened two-particle interaction, which is the second central message of this work. For two arbitrary particles with center-to-center distance vector ${\bf d}$ and orientations $\Omega_{1,2}$ that are mapped to charge distributions $q_i({\bf r};{\bf d},\Omega_i)$ $(i=1,2)$, respectively, I find for the effective pair interaction within the DH approximation
\begin{equation}
\beta \Phi_e({\bf d},\Omega_1,\Omega_2)=\ell_B\int d{\bf r}'q_1({\bf r})q_2({\bf r}')\frac{\exp\left(-\kappa|{\bf r}-{\bf r}'|\right)}{|{\bf r}-{\bf r}'|},
\label{eq:enDH}
\end{equation}
where I subtracted the infinite self-energy terms, and for simplicity I omitted the explicit configurational dependence of the charge distributions in the right-hand side of the equation.

\begin{figure*}[t]
\centering
\includegraphics[width=\textwidth]{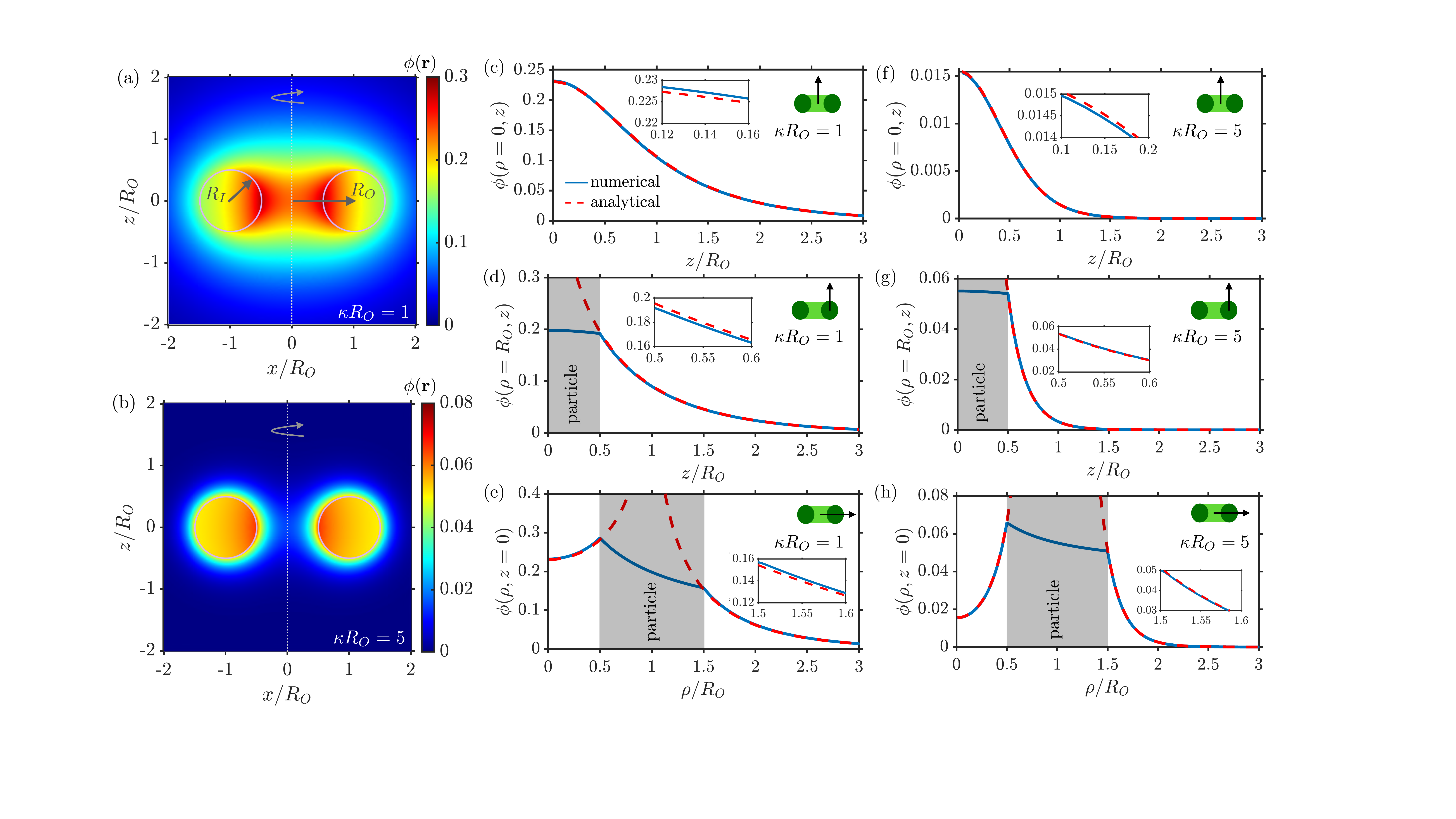}
\caption{Electrostatic potential $\phi({\bf r})/(\beta e)$ around a charged torus with charge $Z_te=50e$, $\epsilon_p/\epsilon_s=0.2$, $\ell_B/R_O=0.01$, and $R_I/R_O=0.5$ for various $\kappa R_O$. For $\kappa R_O=1$ ($\kappa R_O=5$), I found $\Upsilon_t=1.22$ ($\Upsilon_t=5.409$) and $R/R_O=0.9863$ ($R/R_O=1.027$). (a) Numerical results for thick and (b) thin double layers. (c)-(h) Comparison of the semi-analytical ring-charge mapping Eq. \eqref{eq:ringhere} with finite-element calculations of the Poisson-Boltzmann equation for thin and thick double layers, along various axes (insets), as well as an enlarged comparison (near the particle surface). }
\label{fig:toruspot}
\end{figure*}

\section{Benchmarks of the theory}
\subsection{Charged spheres}
As an example, a sphere of radius $a$ and total charge $Z_se=4\pi a^2\sigma e$ can be mapped to a point charge $q({\bf r})=Q_p\delta({\bf r})$. 
Using Eq. \eqref{eq:DHconv} gives $\varphi({\bf r})=Q_p\ell_B\exp(-\kappa r)/{r}$, to be compared with the exact analytical solution \cite{Bell:1970}
\begin{gather}
\phi_\mathrm{>}({\bf r})=\frac{Z_s\ell_B\exp(\kappa a)}{1+\kappa a}\frac{\exp(-\kappa r)}{r}.
\label{eq:sphereout}
\end{gather}
Matching the potential on the particle surface, \mbox{${\varphi}(r=a)={\phi}_\mathrm{>}(r=a)$}, I find $Q_p=Z_s\Upsilon_s$, with $\Upsilon_s=\exp(\kappa a)/(1+\kappa a)$. Alternatively, $Q_p$ can be computed: For the ion densities caused by the singular point charge ${\rho}_\pm({\bf r})$, one can check that $Z_s=Q_p+\int_{r<a}d{\bf r}[{\rho}_+({\bf r})-{\rho}_-({\bf r})]$, showing that an ion-impenetrable charged sphere produces the same electrostatic potential for $r>a$ as an ion-penetrable particle, consisting of a suitable point charge surrounded by a plasma of ions. This physical interpretation is possible because $\phi_\mathrm{<}({\bf r})=Z_s\ell_B/[a(1+\kappa a)]$ is constant and therefore does not contribute to Eq. \eqref{eq:BC}. 

For spherical particles, the DLVO potential results from Eq. \eqref{eq:enDH},
\begin{equation}
\beta\Phi_e^{ss}(d)=Z_s^2\ell_B\left[\frac{\exp(\kappa a)}{1+\kappa a}\right]^2\frac{\exp(-\kappa d)}{d},
\label{eq:DLVO}
\end{equation}
with $d=|{\bf d}|$. The same result follows from the linear superposition approximation (LSA) used on Eq. \eqref{eq:sphereout}, $\phi_{2B}({\bf r})\approx\sum_i\phi_\mathrm{>}({\bf r}-{\bf X}_i)$, in force calculations with the stress tensor \cite{Bell:1970}. However, the free energy route as in Eq. \eqref{eq:enDH} is incompatible with the LSA: One would obtain a different result than Eq. \eqref{eq:DLVO} because the LSA does not account for ion-particle hard-core interactions \cite{Trizac:2002}. However, unique for the point-charge mapping is the compatibility with the free energy \emph{and} the stress tensor route, because the particle hard core is effectively mapped out. Finally, the choice of $q({\bf r})$ is not unique: Mapping spheres to ion-penetrable charged spherical shells gives the same result as the point-charge mapping.

\begin{figure*}[t]
\centering
\includegraphics[width=\textwidth]{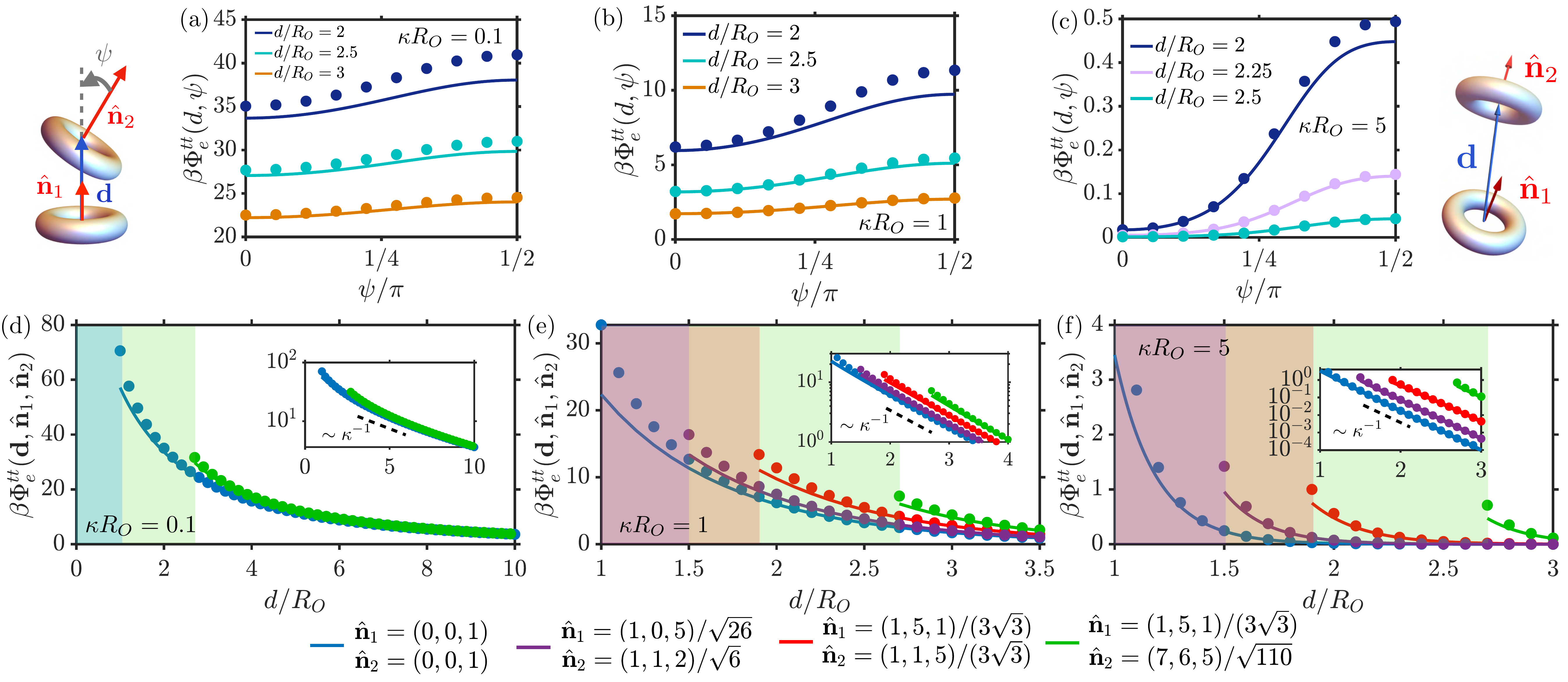}
\caption{Semianalytical approximation of the effective pair potential for identical tori [Eq. \eqref{eq:tt}, lines] compared with finite-element calculations (dots) for various particle configurations and screening lengths at  $\ell_B/R_O=0.01$ and $\epsilon_p/\epsilon_s=0.2$. The coloured shaded areas indicate overlap of particle hard cores when the particle distance $d$ is decreased at fixed orientation from $d\rightarrow\infty$. Tori have charge $Z_t=100$ and inner radius $R_I=0.5R_O$. }
\label{fig:intpotcompare}
\end{figure*}

\subsection{Charged tori}
Next, I show an example where the particle is mapped to a line charge $\mathcal{C}$. Such a mapping is akin to the slender-body theory for creeping flow \cite{Keller:1976}; however, my method works also for ``thick" particles, because $\mathcal{C}$ need not coincide with the centerline of the particle. As an illustration, I consider the pair potential between two identical tori with inner radius $R_I$, outer radius $R_O$ [Fig. \ref{fig:toruspot}(a)], and uniform surface charge density $\sigma=Z_t/(4\pi^2R_IR_O)$. As a first step, I map the torus to a charged ring  $\mathcal{C}$ with parametrisation $\boldsymbol{\gamma}: [0,2\pi)\rightarrow\mathcal{C}$ given by $\boldsymbol{\gamma}(u)=(R\cos u, R\sin u,0)$, of uniform line charge density $\lambda=Q_r/(2\pi R)$. Unlike the point-charge mapping, not only does the line charge number $Q_r$ have to be determined, but also the shape for \mbox{$R\in[R_O-R_I,R_O+R_I]$}. Using Eq. \eqref{eq:DHconv}, I find,
\begin{equation}
\varphi({\bf r})=\frac{Z_t\ell_B\Upsilon_t(\kappa R_O,R_I/R_O)}{2\pi}\int_0^{2\pi}du\frac{\exp\left[-\kappa|{\bf r}-\boldsymbol{\gamma}(u)|\right]}{|{\bf r}-\boldsymbol{\gamma}(u)|},
\label{eq:ringhere}
\end{equation}
where I factorised the ring charge $Q_r=Z_t\Upsilon_t$ using the linearity of Eq. \eqref{eq:DH}.
Note that only for $\kappa=0$ (and along the $z$ axis) can the integral in Eq. \eqref{eq:ringhere} be evaluated in terms of known special functions \cite{Hafeez:2009}. For the general case the integral can be numerically computed on a desktop PC within seconds. Second, to establish the values of $\Upsilon_t$ and $R$, I fit the surface potential $\varphi(\rho=R_O+R_I\cos\alpha,z=R_I\sin\alpha)$, with $\alpha\in[0,2\pi)$, to the numerically obtained surface potential of a charged torus for fixed $\kappa R_O$, $R_I/R_O$, and $\epsilon_p/\epsilon_s$; see Appendix B for details.

For the numerically obtained axisymmetric $\phi({\bf r})$ of a torus [Figs. \ref{fig:toruspot}(a-b)] \cite{Everts:2018}, I show how the semianalytical approximation compares for thin and thick double layers with finite-element calculations (see also Appendix C) for various cuts along the torus [Figs. \ref{fig:toruspot}(c-h)]. For weak and strong intraparticle double-layer overlap, Eq. \eqref{eq:ringhere} agrees excellently with numerics, capturing the full spatial dependence of $\phi_>({\bf r})$ for suitably chosen $\Upsilon$ and $R$, even for tori with large particle volumes. Finally, Eq. \eqref{eq:ringhere} gives analytical insight: For $r\rightarrow\infty$, I find that $\varphi({\bf r})\sim\mathcal{A}(\theta)Z_t\ell_B\exp(-\kappa r)/r$, with anisotropy function (see Appendix D for the derivation),
\begin{equation}
\frac{\mathcal{A}(\theta)}{\Upsilon_t}=1+\frac{1}{4}\sin^2\theta\,(\kappa R)^2+\frac{1}{64}\sin^4\theta\,(\kappa R)^4+...
\label{eq:anisofunc}
\end{equation}
showing the well-known result that particle anisotropy still persists in the far-field electrostatic potential unlike the unscreened case, and that the anisotropies are more pronounced for large $\kappa$ \cite{Trizac:2002, Tellez:2010}. To my best knowledge $\mathcal{A}(\theta)$ has never been calculated for a torus before.

\begin{figure*}[t]
\centering
\includegraphics[width=\textwidth]{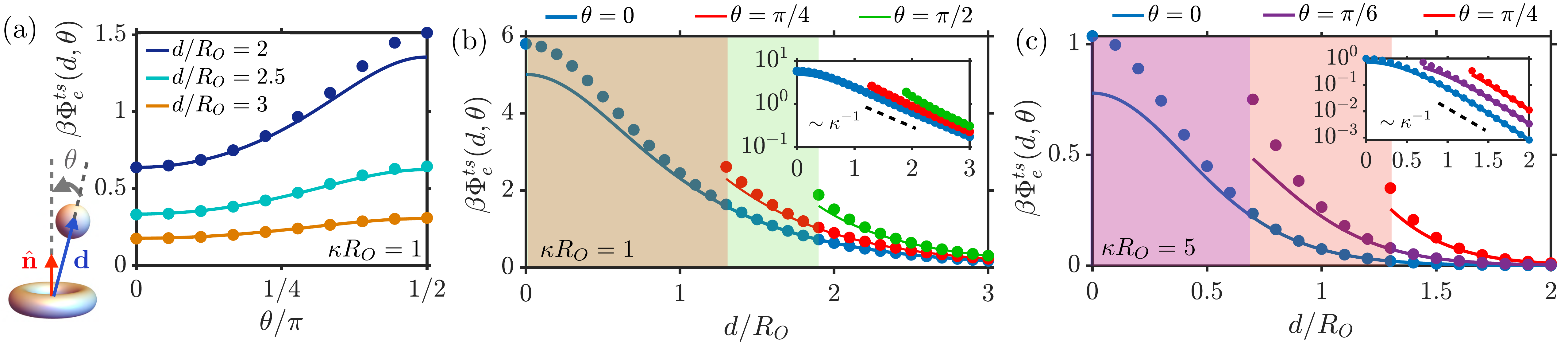}
\caption{Comparisons of the semianalytical approximation of the effective pair potential (lines) with numerical finite-element calculations (dots) between a charged torus and a charged sphere [Eq. \eqref{eq:ts}] with arbitrarily chosen particle configurations and screening constants. The coloured shaded areas indicate forbidden regions due to the particle hard cores when the particle distance $d$ is decreased at fixed orientation from $d\rightarrow\infty$. The tori have charge $Z_t=100$ and inner radius $R_I=0.5R_O$. Spheres have the same surface charge density as the torus with radius $a=0.4R_O$. For all cases $\ell_B/R_O=0.01$ and $\epsilon_p/\epsilon_s=0.2$.}
\label{fig:ts}
\end{figure*}

To calculate effective pair interactions, I parametrise two identical rings with arbitrary orientations as $\boldsymbol{\gamma}_i(u)={\bf X}_i+R\cos u \hat{\bf l}_i+R\sin u\hat{\bf m}_i$,  $(i=1,2)$,
with $\hat{\bf n}_i\cdot\hat{\bf l}_i=\hat{\bf n}_i\cdot\hat{\bf m}_i=0$, and $\hat{\bf n}_i$ defined in the inset of Fig. \ref{fig:intpotcompare}(b). Using Eq. \eqref{eq:enDH}, I find the interaction between two identical tori
\begin{align}
\beta\Phi_e^{tt}&({\bf d},\hat{\bf n}_1,\hat{\bf n}_2)=\frac{Z_t^2\ell_B}{(2\pi)^2}\Upsilon_t(\kappa R_O,\kappa R_I)^2 \nonumber\\
&\times\int_0^{2\pi}du \int_{0}^{2\pi}dv \, \frac{\exp{[-\kappa|\boldsymbol{\gamma}_1(u)-\boldsymbol{\gamma}_2(v)|]}}{|\boldsymbol{\gamma}_1(u)-\boldsymbol{\gamma}_2(v)|}. \label{eq:ttm}
\end{align}
Note that the difference $|\boldsymbol{\gamma}_1(u)-\boldsymbol{\gamma}_2(v)|$ depends on ${\bf d},\hat{\bf n}_1,$ and $\hat{\bf n}_2$.  Despite being an integral representation, the evaluation of Eq. \eqref{eq:ttm} is far less computationally expensive than three-dimensional finite-element calculations of Eq. \eqref{eq:DH} for every fixed particle configuration. Explicit calculations give
\begin{widetext}
\begin{align}
\beta & \Phi_e^{tt}({\bf d},\hat{\bf n}_{1},\hat{\bf n}_{2})=\frac{Z_t^2\ell_B\Upsilon_t^2}{(2\pi)^2}\int_0^{2\pi}du\int_0^{2\pi}dv \ \frac{\exp{\left[-\kappa\sqrt{d^2+2R^2-2Rf(u,v;{\bf d},\hat{\bf n}_{1},\hat{\bf n}_{2})}\right]}}{\sqrt{d^2+2R^2-2Rf(u,v;{\bf d},\hat{\bf n}_{1},\hat{\bf n}_{2})}}, \label{eq:tt}
\end{align}
for ${\bf d}$ not parallel to $\hat{\bf n}_1$ and/or $\hat{\bf n}_2$, where the term $f$ in the integrand simplifies to
\begin{align}
f(u,v; &\ {\bf d},\hat{\bf n}_{1},\hat{\bf n}_{2})=d\sin v|\hat{\bf n}_1\times\hat{\bf n}_2| +R\left[\cos(u-v)-(1-\hat{\bf n}_1\cdot\hat{\bf n}_2)\sin u\sin v\right], 
\end{align}
whereas for ${\bf d}$ not parallel to either $\hat{\bf n}_1$ or $\hat{\bf n}_2$, 
\begin{align}
f&=|{\bf b}_1|\sin u-|{\bf b}_2|\sin v+\frac{R}{|{\bf b}_1||{\bf b}_2|}\Big(\cos u \cos v\left[d^2(\hat{\bf n}_1\cdot\hat{\bf n}_2)-({\bf d}\cdot\hat{\bf n}_1)({\bf d}\cdot\hat{\bf n}_2)\right]+\cos u\sin v\left[{\bf d}\cdot(\hat{\bf n}_1\times\hat{\bf n}_2)({\bf d}\cdot\hat{\bf n}_2)\right] \nonumber \\
&-\sin u\cos v\left[{\bf d}\cdot(\hat{\bf n}_1\times\hat{\bf n}_2)({\bf d}\cdot\hat{\bf n}_1)\right]+\sin u\sin v\left\{d^2(\hat{\bf n}_1\cdot\hat{\bf n}_2)^2-({\bf d}\cdot\hat{\bf n}_1)({\bf d}\cdot\hat{\bf n}_2)(\hat{\bf n}_1\cdot\hat{\bf n}_2)+\left[{\bf d}\cdot(\hat{\bf n}_1\times\hat{\bf n}_2)\right]^2\right\}\Big). \label{eq:bigf}
\end{align}
\end{widetext}
In Eq. \eqref{eq:bigf}, I used the parametrisation $\hat{\bf l}_i={\bf b}_i/|{\bf b}_i|$ and $\hat{\bf m}_i=\hat{\bf l}_i\times\hat{\bf n}_i$, with vector ${\bf b}_i={\bf d}\times\hat{\bf n}_i$ for $i=1,2$.

In Fig. \ref{fig:intpotcompare}, I compare the numerical calculation and the analytical approximation of Eqs. \eqref{eq:tt} for a wide variety of particle configurations and Debye screening lengths. I find an excellent agreement with only a deviation for particle separations close to contact where the repulsion is underestimated, as expected from a method that is equivalent to the LSA  \cite{Carnie:1993}. 

%In Fig. 2 of the main text I compared numerically obtained torus-torus interactions with Eqs. (10-12) for particularly chosen particle separations and orientations. The comparisons were only performed at $\kappa R_O=1$ which is a double layer with similar thickness as the particle size, and consequently strong intra-particle double layer overlap. I add to this comparison a case where the double layer is much larger than the particle size ($\kappa R_O=0.1$) and much thinner than the particle size ($\kappa R_O=5$), and hence weak intra-particle double layer overlap. For the first case, from Fig. \ref{fig:fits} I find $\Delta=1.0029$ and $R=0.9436$. The comparisons are shown in Fig. \ref{fig:SIpairpot} showing again excellent agreement with weaker anistropy for $\kappa R_O=0.1$ compared to $\kappa R_O=5$ as explained in the main text. For clarity, purple and red lines are not plotted in Fig. \ref{fig:SIpairpot}(b), because of the too weak anisotropy.

\subsection{Additive torus-sphere mixtures}
The method can also be extended to additive mixtures of particles, as I highlight for the torus-sphere interaction,
\begin{align}
\beta&\Phi_e^{ts}({\bf d},\hat{\bf n})=\frac{Z_tZ_s\ell_B}{2\pi}\Upsilon_t(\kappa R_O,\kappa R_I)\Upsilon_s(\kappa a) \nonumber \\
&\times\int_0^{2\pi}du\, \frac{\exp{\left[-\kappa\sqrt{d^2+R^2+2R\sin u|{\bf d}\times\hat{\bf n}|}\right]}}{\sqrt{d^2+R^2+2R\sin u|{\bf d}\times\hat{\bf n}|}},
\label{eq:ts}
\end{align}
which, again, agrees excellently with numerics [Fig. \ref{fig:ts}], even when the sphere partly enters the hole of the torus [Fig. \ref{fig:ts}(b,c), blue line]. In the insets, I highlight on a log-linear scale that the decay length for these particle shapes is still $\kappa^{-1}$; however, with an orientation-dependent interaction amplitude of higher anisotropy when the salt concentration is increased [Figs. \ref{fig:ts}(b,c)], which is generic for anisotropic particles \cite{Trizac:2002, Tellez:2010}, as is for the electrostatic potential [Eq. \eqref{eq:anisofunc}].

\begin{figure*}[t]
\centering
\includegraphics[width=0.95\textwidth]{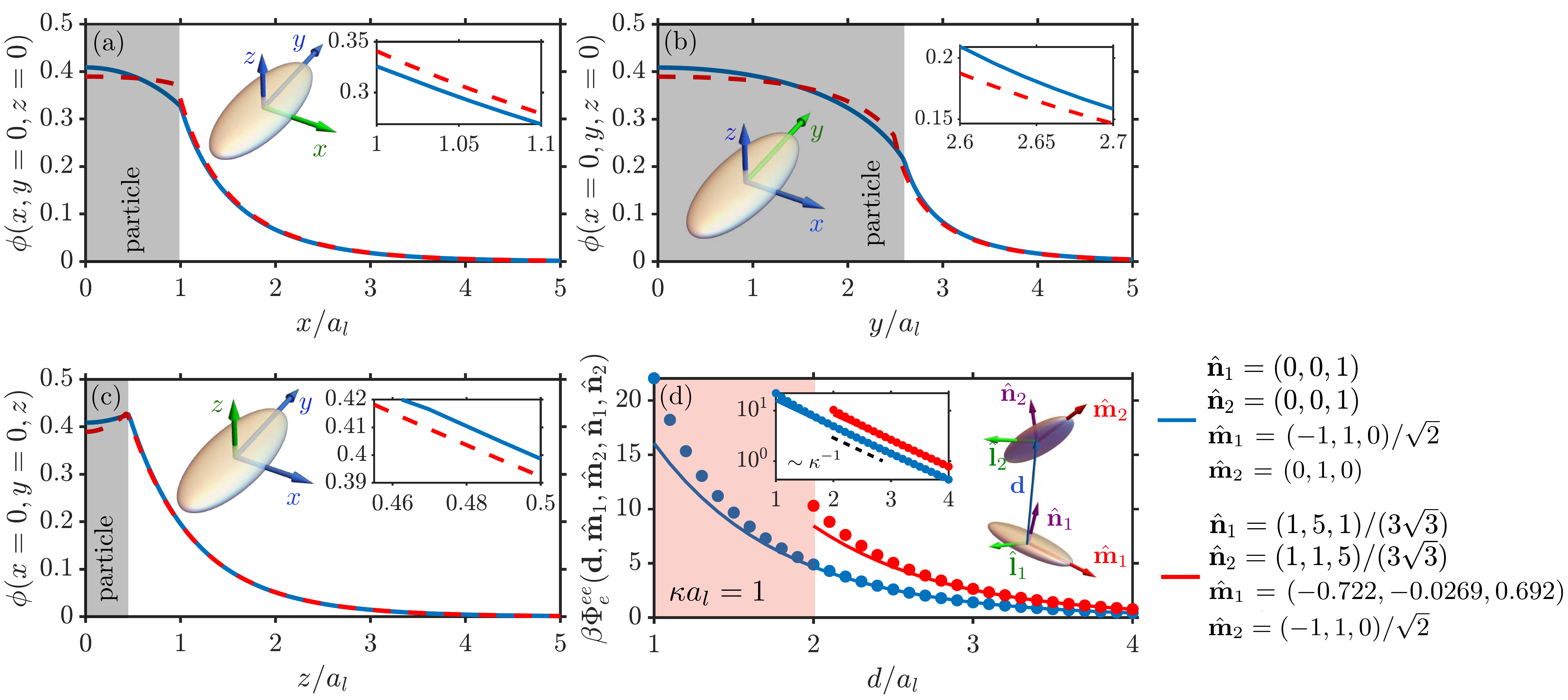}
\caption{Comparison of (a-c) electrostatic potential [Eq. \eqref{eq:pote}] and (d) the effective pair potential [Eq. \eqref{eq:eee}] of the semianalytical ellisoidal-shell mapping with numerical finite-element calculations of the Poisson-Boltzmann equation for a triaxial ellipsoidal particle at various values of $\kappa a_l$, with $a_l$ being the distance from the origin to the particle surface along the $x$ axis. The potential is compared along various cuts around the ellipsoid, as well as an enlargement of the comparison between numerical and analytical results close to the particle surface. In all plots, $\epsilon_p/\epsilon_s=0.2$, $\ell_B/a_l=0.01$ and $Z=100$. The shell that has been used to represent the particle has the same ratios of the ellipsoidal main axes, but with a rescalable parameter $R_l<a_l$ and charge parameter $\Upsilon$ with the same meaning as for the torus. For precise particle dimensions, see the main text.}
\label{fig:ellipsoidpot}
\end{figure*}

\subsection{Charged tri-axial ellipsoids}
To show the generality of the framework, I consider the interaction potentials of two triaxial ellipsoids, for which a mapping to a spherical shell is needed, rather than a line charge. I hypothesise, however, that some metal spheroidal particles can be mapped to straight lines, because the (unscreened) isopotential surfaces of straight lines are prolate spheroids with specific aspect ratios \cite{Curtright:2016}. It is straightforward to generalise the approach to mappings to ion-penetrable charged surfaces $\mathcal{S}$ with surface charge distribution $\sigma_s$, and $\mathcal{S}$ parametrised by $\boldsymbol{\Gamma}: [a,b]\times[c,d]\rightarrow\mathcal{S}$. I find from Eq. (5)
\begin{align}
\varphi({\bf r})=\ell_B\int_a^bdu\int_{c}^{d}dv\,&\sqrt{g(u,v)}\ \sigma_s(\boldsymbol{\Gamma}(u,v)) \nonumber \\
&\times\frac{\exp[-\kappa|{\bf r}-\boldsymbol{\Gamma}(u,v)|]}{|{\bf r}-\boldsymbol{\Gamma}(u,v)|}, \label{eq:surf}
\end{align}
where $g(u,v)$ is the determinant of the induced metric tensor.

For the particle, consider the parametrisation for a triaxial ellipsoid
\begin{equation}
{\bf X}(\phi,\theta)=(a_l\cos \phi\sin \theta,a_m\sin \phi \sin \theta ,a_n\cos \theta),
\end{equation}
where the angles are defined for $\phi\in[0,2\pi)$ and $\theta\in[0,\pi)$, with volume ${V_{e}=(4/3)\pi a_l a_m a_n}$,
and surface area
\begin{align}
S_{e}&(a_l,a_m,a_n)=2\pi a_n^2 \nonumber \\
&+\frac{2\pi a_l a_m}{\sin\psi}\left[E(\psi,k)\sin^2\psi+F(\psi,k)\cos^2\psi\right], 
\end{align}
with $F(\psi,k)$ and $E(\psi,k)$ being the incomplete elliptic integrals of the first and second kinds, respectively. Furthermore, $\cos\psi=a_n/a_l$ and $k^2=a_l^2/a_m^2(a_m^2-a_n^2)/(a_l^2-a_n^2)$.
Since ${\bf X}$ and ${\bf \Gamma}$ should have the same shape, I use an ellipsoidal shell with the same aspect ratios as the particle, but with a smaller size parameterised by $R_l$,
\begin{equation}
{\bf \Gamma}(u,v)=R_l(\cos u\sin v,\gamma_m\sin u \sin v ,\gamma_n\cos v), 
\end{equation}
with $u\in[0,2\pi)$, $v\in[0,\pi)$, $\gamma_m=a_m/a_l$, and $\gamma_n=a_n/a_l$.

From Eq. \eqref{eq:surf}, I determine the electrostatic potential of a tri-axial ellipsoidal particle
\begin{widetext}
\begin{align}
\varphi({\bf r})=\frac{Z_e\ell_B\Upsilon R_l^2}{S_e(R_l,\gamma_m R_l, \gamma_n R_l)}&\int_0^{2\pi}  du\int_0^\pi dv\, \sqrt{\gamma_m^2\cos^2v\sin^2v+\gamma_n^2\sin^4v\left(\gamma_m^2\cos^2u+\sin^2u\right)} \nonumber \\
&\times\frac{\exp\left[-\kappa\sqrt{\left(x-R_l\cos u \sin v\right)^2+\left(y-\gamma_m R_l\sin u \sin v\right)^2+\left(z-\gamma_n R_l\cos v\right)^2}\right]}{\sqrt{\left(x-R_l\cos u \sin v\right)^2+\left(y-\gamma_m R_l\sin u \sin v\right)^2+\left(z-\gamma_n R_l\cos v\right)^2}},
\label{eq:pote}
\end{align}
\end{widetext}
where I used for $\sigma_s$ a homogeneously charged ellipsoidal shell with total charge  $Z_e\Upsilon e$. The parameters $R_l$ and $\Upsilon$ are determined via a fit of the surface potential $\varphi_0(\phi,\theta)=\varphi({\bf X}(\phi,\theta))$. 
As an example, I determine $\Upsilon$ and $R_l$ for a tri-axial ellipsoid with the same volume and surface area as a torus with inner radius $R_I=0.5 R_O$ and $a_l=R_O$. In this case the aspect ratios are $\gamma_m=2.595$ and $\gamma_n=0.454$. Moreover, I fix $\kappa a_l=1$ and $\epsilon_p/\epsilon_s=0.2$. The result of the fit is $R_l=0.9599$ and $\Upsilon=1.1523$. In Figs. \ref{fig:ellipsoidpot}(a-c) I show a comparison of the resulting electrostatic potential with finite-element calculations, showing excellent agreement outside the particle. 

\begin{widetext}
Finally, by using Eq. \eqref{eq:enDH} I derive for the effective pair interaction between two particles that are mapped to charged shells,
\begin{align}
\beta\Phi_e=\ell_B&\int_{a_1}^{b_1}du\, \int_{c_1}^{d_1}dv\, \sqrt{g_1(u,v)}\sigma_{s,1}(\boldsymbol{\Gamma}_1(u,v)) \int_{a_2}^{b_2}du'\int_{c_2}^{d_2}dv' \sqrt{g_2(u',v')}\sigma_{s,2}(\boldsymbol{\Gamma}_2(u',v'))\frac{\exp{[-\kappa|\boldsymbol{\Gamma}_1(u,v)-\boldsymbol{\Gamma}_2(u',v')|]}}{|\boldsymbol{\Gamma}_1(u,v)-\boldsymbol{\Gamma}_2(u',v')|}.  \label{eq:s}
\end{align}
I apply Eq. \eqref{eq:s} to arbitrarily oriented and positioned ellipsoidal particles, mapped to (for $i=1,2$)
\begin{equation}
\boldsymbol{\Gamma}_i(u,v)=R_l\cos u\sin v\,\hat{\bf l}_i+R_m\sin u\sin v\,\hat{\bf m}_i+R_n\cos v\,\hat{\bf n}_i, \quad 0\leq u< 2\pi, \quad 0\leq v<\pi.
\label{eq:ppp}
\end{equation}
Here $\{\hat{\bf l}_i,\hat{\bf m}_i,\hat{\bf n}_i\}$ is an orthonormal triad of vectors. I find applying Eq. \eqref{eq:ppp} to Eq. \eqref{eq:s} that
\begin{align}
&\beta\Phi_e^{ee}\left({\bf d},\{\hat{\bf n}_i,\hat{\bf m}_i\}_{i=1}^2\right)=\frac{Z_e^2\ell_B\Upsilon^2R_l^4}{S_e(R_l,\gamma_m R_l, \gamma_n R_l)^2}\int_0^{2\pi}du\int_0^\pi dv\, \sqrt{\gamma_m^2\cos^2v\sin^2v+\gamma_n^2\sin^4v\left(\gamma_m^2\cos^2u+\sin^2u\right)}
 \nonumber \\
&\int_0^{2\pi}du'\int_0^\pi dv'\, \sqrt{\gamma_m^2\cos^2v'\sin^2v'+\gamma_n^2\sin^4v'\left(\gamma_m^2\cos^2u'+\sin^2u'\right)}
\frac{\exp\left[-\kappa\sqrt{f(u,u',v,v',{\bf d},\{\hat{\bf n}_i,\hat{\bf m}_i\}_{i=1}^2)}\right]}{\sqrt{f(u,u',v,v',{\bf d},\{\hat{\bf n}_i,\hat{\bf m}_i\}_{i=1}^2)}}, \label{eq:eee}
\end{align}
where the function $f$ in the integrand is expressed as
\begin{align}
f&(u,u',v,v',{\bf d},\{\hat{\bf n}_i,\hat{\bf m}_i\}_{i=1}^2)=d^2+R_l^2\Big[\cos^2u\sin^2v+\cos^2u'\sin^2v'-2\hat{\bf l}_1\cdot\hat{\bf l}_2\cos u\cos u'\sin v\sin v' \nonumber\\
&-2\gamma_n(\hat{\bf l}_1\cdot\hat{\bf n}_2\cos u\sin v\cos v'+\hat{\bf l}_2\cdot\hat{\bf n}_1\cos u'\sin v'\cos v)-2\gamma_m\gamma_n(\hat{\bf m}_1\cdot\hat{\bf n}_2\sin u\sin v\cos v'+\hat{\bf m}_2\cdot\hat{\bf n}_1\sin u'\sin v'\cos v) \nonumber \\
&-2\gamma_m(\hat{\bf l}_1\cdot\hat{\bf m}_2\sin u'\sin v'\cos u \sin v+\hat{\bf l}_2\cdot\hat{\bf m}_1\sin u\sin v\cos u' \sin v')+\gamma_n^2(\cos^2v+\cos^2 v'-2\hat{\bf n}_1\cdot\hat{\bf n}_2\cos v\cos v') \nonumber \\
&+ \gamma_m^2(\sin^2u\sin^2v+\sin^2u'\sin^2 v'-2\hat{\bf m}_1\cdot\hat{\bf m}_2\sin u\sin u'\sin v\sin v')\Big]+2R_l\Big[{\bf d}\cdot\hat{\bf l}_1\cos u\sin v-{\bf d}\cdot\hat{\bf l}_2\cos u'\sin v' \nonumber \\
&+\gamma_m({\bf d}\cdot\hat{\bf m}_1\sin u\sin v - {\bf d}\cdot\hat{\bf m}_2\sin u'\sin v')+\gamma_n\left({\bf d}\cdot\hat{\bf n}_1\cos v-{\bf d}\cdot\hat{\bf n}_2\cos v'\right)\Big],
\end{align}
\end{widetext}
with $\hat{\bf l}_i=\hat{\bf m}_i\times\hat{\bf n}_i$. This ellipsoid-ellipsoid interaction potential is tested in Fig. \ref{fig:ellipsoidpot}(d) to numerical calculations, showing excellent agreement, except for small particle separations, as was also found in the previous examples.

\section{Conclusions and outlook}
In summary, I developed a framework to derive effective pair potentials between finite-sized arbitrarily shaped rigid macroions. With different combinations of spheres, ellipsoids, and tori in various mutual orientations, I showed the applicability and accuracy of this method: This framework gives analytical insights with broad relevance for experiments and simulations. I expect that the method applies to many shapes when multiple inhomogeneous singular charge distributions are used. Furthermore, as recently shown, the method works also for spheres dispersed in nematic liquid crystals \cite{Everts:2020}. Finding the correct charge distribution for a specific particle can be nontrivial, but often symmetry arguments and an analysis of the necessary multipole moments to match the far-field electrostatic potential are useful considerations. For example, it was suggested that Janus spheres can be described by a collection of point charges \cite{Graaf:2012}. My method becomes less accurate at small particle separations, like DLVO theory, where the surface potential and/or surface charge density become ``polarised" which, however, can be reconciled with the Derjaguin approximation \cite{Bell:1970, Schiller:2011, Oettel:2014, Levin:2019}. Furthermore, it is not possible to derive multibody interactions (of the form as in Ref. \cite{Bechinger:2004}), because of the underlying LSA equivalence.  

As an outlook, I propose extending the theory with (many-body) charge regulation \cite{Grunberg:1999, Everts:2016, Everts:2016b, Markovich:2016, Levin:2019} and renormalisation \cite{Alexander:1984, Levin:2004, *Levin:2007, Pincus:2015, Boon:2015} to incorporate more types of electrostatic boundary conditions and nonlinear screening, respectively. The expressions of this paper can then still be used with the bare charge replaced by an effective (renormalised) charge. Another extension of the theory would be to include correlations beyond the mean-field result presented here. It would be interesting to see in this case whether the mapping to a singular charge distribution is still suitable to describe the (thermally averaged) electrostatic potential and pair interactions accurately. Such questions can be answered via the field-theoretical formulation \cite{Netz:2000} of two particles immersed in an ion-containing solvent. Furthermore, such a field-theoretical formulation would be interesting to investigate the effects of flexible particles, rather than the rigid particles discussed here. In this case, an elastic free-energy contribution would self-consistently determine the singular charge distribution needed for the mapping, and this deserves further research.

 More broadly, the findings might also be valuable for any physical system governed by the Helmholtz equation, e.g. acoustics \cite{Miloh:2016} and optics \cite{Feynman:2011}, or systems where the Yukawa potential is involved, such as wetting \cite{Evans:1983}. Finally, it would be intriguing to explore charge-screened active matter for various ``thick" particles. Here, the Yukawa potential is already often used to model steric repulsions between thin active rods \cite{Binder:2012, Lowen:2012}.

\begin{acknowledgments} I acknowledge financial support from the European Union's Horizon 2020 programme under the Marie Skłodowska-Curie Grant Agreement No. 795377 and the Slovenian Research Agency ARRS under Contract No. J1-9149. Furthermore, I benefited from fruitful discussions with M. Ravnik, M. Murko, S. \v Copar, \mbox{D. J. Lee}, and R. Goldstein. Special thanks goes to \mbox{N. Boon}, who exposed me to the derivation of the DLVO potential using a point-charge mapping, which was an incentive to generalise this method to more complicated structures. \mbox{M. A. Janssen},  S. \v Copar, and C. Schaefer are thanked for critically reading the manuscript and for providing useful comments. Finally, I would like to thank the Isaac Newton Institute for Mathematical Sciences for support and hospitality during the programme [The Mathematical Design of New Materials] when work on this paper was undertaken. This work was supported by: EPSRC Grant No. EP/R014604/1.
\end{acknowledgments}  

\bibliography{literature1} % Tell bibtex which .bib file to use (this one is some example file in TexLive's file tree)

\onecolumngrid

\section*{Appendix A: Numerical details}
\renewcommand{\theequation}{A.\arabic{equation}}
\renewcommand{\thefigure}{A\arabic{figure}}
\setcounter{figure}{0}
\setcounter{equation}{0}

In the main text, I compared analytical expressions within linear screening theory with numerical calculations of the electrostatic potential and the effective pair interactions. In all cases, the comparisons were made using the non-linear Poisson-Boltzmann equation,
\begin{gather}
\nabla^2\phi_<({\bf r})=0, \quad {\bf r}\in\mathrm{int}(\mathcal{P}), \\
\nabla^2\phi_>({\bf r})=\kappa^2\sinh[\phi({\bf r})], \quad {\bf r}\notin\mathrm{int}(\mathcal{P}),
\end{gather}
subjected to the constant-charge boundary conditions Eq. (2) of the main text. These set of differential equations are solved with COMSOL Multiphysics 5.4, and where needed the cylindrical symmetry of the problem has been exploited (in particular for a single torus). For three-dimensional finite-element calculations, boundary layers were used to resolve the double layer close to the particle surface and a sufficiently large system size is used to ensure that $\phi(r\rightarrow\infty)=0$. As mesh I used free tetrahedral elements, using extremely fine elements within the general physics settings. In this work, particle charges are chosen within the linear-screening regime.

For the pair interactions, I integrated using the built-in integration operators the following expression
\begin{align}
\beta & H({\bf d},\Omega_1, \Omega_2)=\beta\Phi_\mathrm{HC}({\bf d},\Omega_1, \Omega_2)+\sum_{i=1}^2\frac{1}{2}\int_{\mathcal{P}_i} d^2{\bf r}\, \sigma\phi({\bf r})+\rho_s\int_\mathcal{R} d^3{\bf r}\, \{\phi({\bf r})\sinh{\phi}({\bf r})-2[\cosh\phi({\bf r})-1]\},
\end{align}
with $\mathcal{R}=V\backslash({\mathrm{int}(\mathcal{P}_1)}\cup {\mathrm{int}(\mathcal{P}_2)})$ the region outside the particles, see for a derivation e.g. Ref. \cite{Everts:2017}. The second term is an entropy term, that in most cases is much smaller than the electrostatic terms (third term). For completeness, I added a particle-particle hard-core interaction $\Phi_\mathrm{HC}({\bf d},\Omega_1, \Omega_2)$ which equals zero when particles do not overlap and is infinity when there is particle overlap. Moreover, I subtract the total self-energy of the two-body system, which I calculate numerically as $H(d\rightarrow\infty,\Omega_1,\Omega_2)$, which is independent of particle orientations $\Omega_{1,2}$, i.e. $\Phi({\bf d},\Omega_1, \Omega_2)= H({\bf d},\Omega_1, \Omega_2)-{H({d}\rightarrow\infty,\Omega_1, \Omega_2)}$.

To numerically evaluate the expressions obtained from a singular-charge distribution mapping, such as Eqs. (9), (11), and (14) from the main text, and Eq. \eqref{eq:pote} in this document, I used the standard \texttt{integral} and \texttt{integral2} commands in MATLAB. For the ellipsoid-ellipsoid interaction Eq. \eqref{eq:eee}, iterative use of these commands was made.

\section*{Appendix B: Determination of charge and shape parameters for an electrostatically screened charged torus}
\renewcommand{\theequation}{B.\arabic{equation}}
\renewcommand{\thefigure}{B\arabic{figure}}
\setcounter{figure}{0}
\setcounter{equation}{0}

\begin{figure*}[t]
\centering
\includegraphics[width=\textwidth]{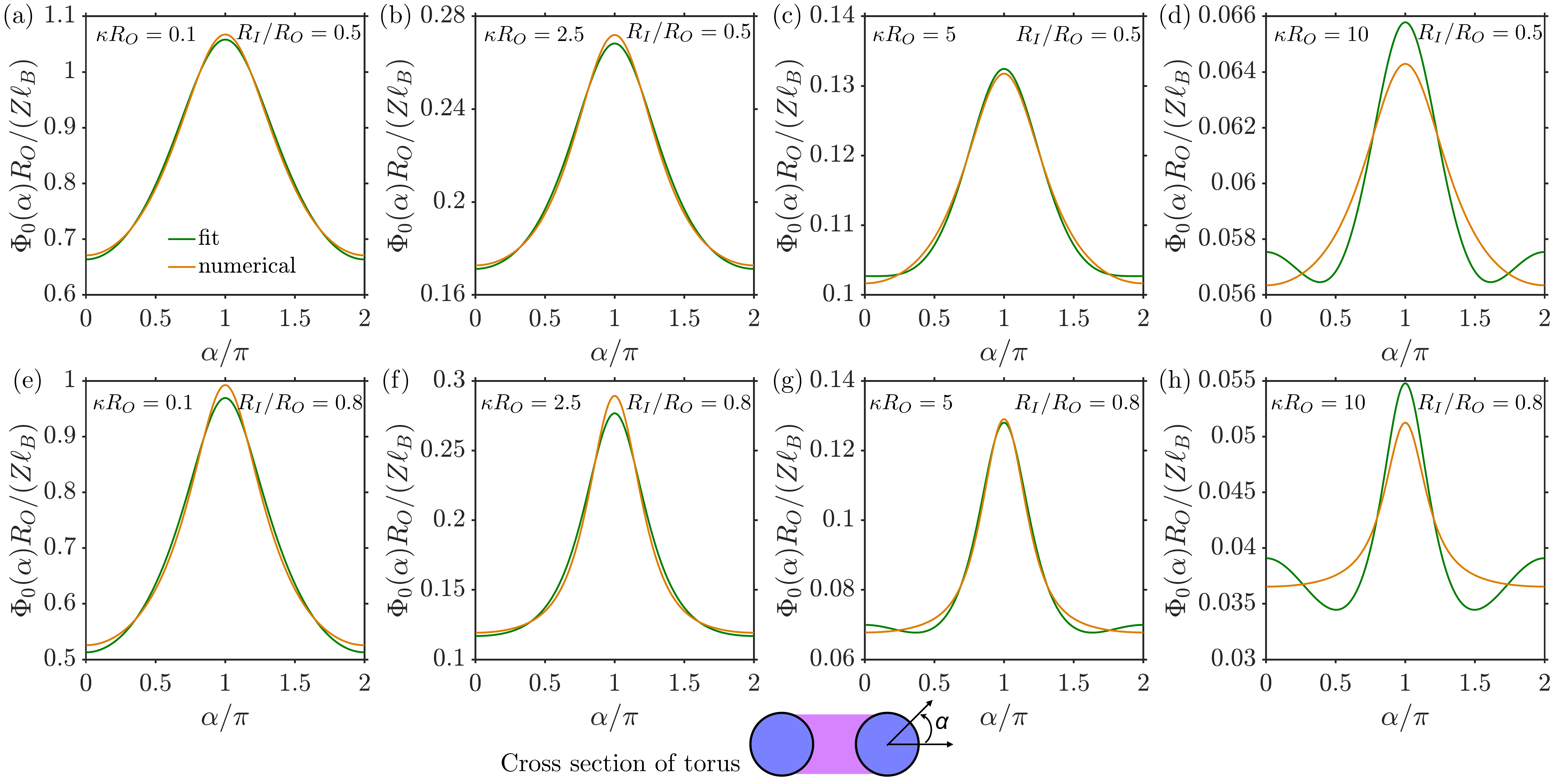}
\caption{Examples of fits of the numerically obtained surface potential (scaled with the Bjerrum length $\ell_B$ and particle charge $Z$) with the expression of a ring charge mapping, Eq. \eqref{eq:ring}. From this fit the values for $R$ and $\Upsilon$ are obtained. All calculations were performed at $\epsilon_p/\epsilon_s=0.2$.}
\label{fig:fits}
\end{figure*}

\begin{figure*}[t]
\centering
\includegraphics[width=\textwidth]{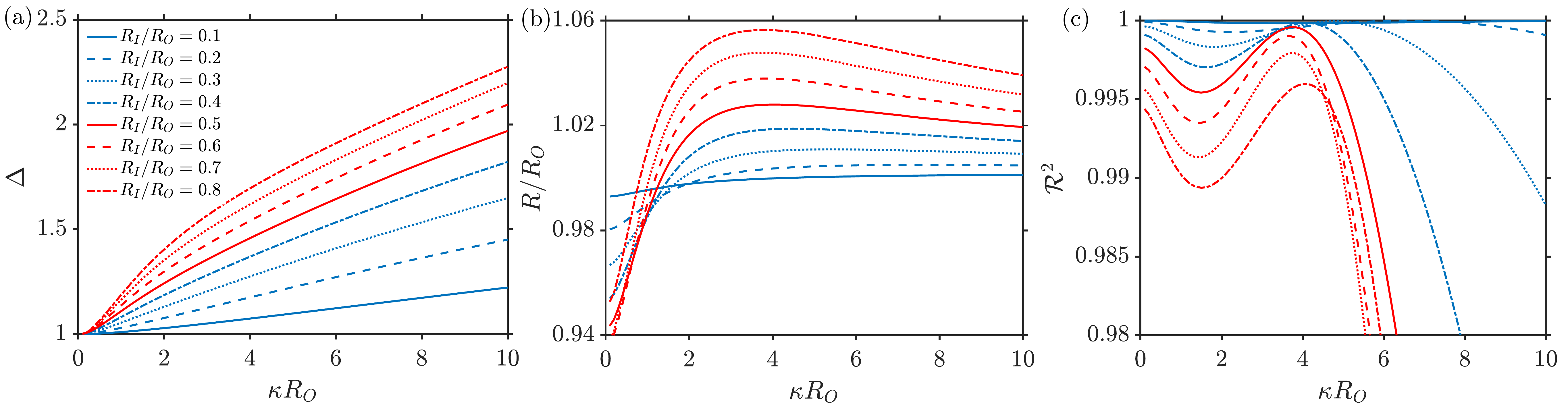}
\caption{Result of (a) the charge parameter $\Delta$ and (b) shape parameter $R$ as obtained from fits of the surface potential, see Fig. \ref{fig:fits}, for various values of $\kappa R_O$ and torus thicknesses as quantified by $R_I/R_O$. The results are independent of particle charge $Z$ and Bjerrum length $\ell_B$ in the linear screening regime, but depend on $\epsilon_p/\epsilon_s$ (fixed at 0.2 as in Fig. \ref{fig:fits}). In  panel (c), we plot a statistical parameter showing the quality of the fits, the so-called coefficient of determination $\mathcal{R}^2$. A value of $\mathcal{R}^2=1$ indicates a perfect fit.}
\label{fig:parameters}
\end{figure*}

Recall the approximation of the electrostatic potential for a single torus Eq. \eqref{eq:ringhere} written down in cylindrical coordinates,
\begin{equation}
\varphi({\bf r})=\frac{Z\ell_B\Upsilon }{2\pi}\int_0^{2\pi}du\, \frac{\exp\left[-\kappa\sqrt{\rho^2+R^2+z^2-2\rho R\cos(u)}\right]}{\sqrt{\rho^2+R^2+z^2-2\rho R\cos(u)}},
\label{eq:ring}
\end{equation}
which for suitably chosen values of $R$ and $\Upsilon$ approximately describes a toroidal particle $\mathcal{P}$ with standard parametrization ${\bf X}(\psi,\alpha)=\big((R_O+R_I\cos\alpha)\cos\psi,(R_O+R_I\cos\alpha)\sin\psi,R_I\sin\alpha\big)$, $\psi,\alpha\in[0,2\pi)$. Consequently, I find the approximation for the surface potential
\begin{equation}
\varphi_0(\alpha)=\frac{ Z\ell_B\Upsilon}{2\pi}\int_0^{2\pi}du\, \frac{\exp\left[-\kappa\sqrt{R^2+R_O^2+R_I^2+2R_OR_I\cos\alpha-2(R_O+R_I\cos\alpha) R\cos(u)}\right]}{\sqrt{R^2+R_O^2+R_I^2+2R_OR_I\cos\alpha-2(R_O+R_I\cos\alpha) R\cos(u)}}.
\label{eq:ring2}
\end{equation}
In MATLAB I use the \texttt{fit} command from the Curve Fitting Toolbox to establish the values of $\Upsilon$ and $R$ by fitting Eq. \eqref{eq:ring2} to numerically obtained surface potentials from COMSOL using the Levenberg-Marquadt algorithm. A few examples of fits are shown in Fig. \ref{fig:fits}. The fits are of good quality even for large particle volumes. The quality of the approximation, however, seems to be the least accurate for extremely thin double layers $\kappa R_O=10$, although this turns out to be of little importance for the accuracy of the full spatial dependence of the electrostatic potential around the particle, as I shall demonstrate in Sec. S3.
It is meaningful to explore the values of the charge parameter $\Upsilon$ and shape parameter $R$ for a wide variety of particle thicknesses as quantified by $R_I/R_O$ and screening lengths as quantified by $\kappa R_O$. For this I fitted a whole set of numerically obtained surface potentials to Eq. \eqref{eq:ring2}. Regarding $\Upsilon$, it is meaningful for the comparison between different $\kappa R_O$ and $R_I/R_O$ to scale out a large factor for better comparison, 
\begin{equation}
\Upsilon(\kappa R_O,R_I/R_O)=\Delta(\kappa R_O,R_I/R_O)\frac{\exp(\kappa R_I)}{1+\kappa R_I},
\end{equation}
and plot $\Delta$ and $R$ as function of $\kappa R_O$ for various $R_I/R_O$ in Figs. \ref{fig:parameters}(a-b).  Observe that $\Delta$ increases with $\kappa R_O$ and $R_I/R_O$, and therefore also $\Upsilon$ increases. For spheres ($\Upsilon=\exp(\kappa a)/(1+\kappa a)$) the same effect is observed for $\Upsilon$ as function of $\kappa a$, however, $\Delta$ (with natural definition) equals unity in this case. Regarding the ring, one can see from Fig. \ref{fig:parameters}(b) that it roughly coincides with the centerline of the torus with nonmonotonous behaviour as function of $\kappa R_O$ and $R_I/R_O$ around $R=R_O$. Although choosing $R=R_O$ would be still a good approximation, as I have checked, the quality of the fit is a little bit better if it is used as a free parameter. The main determining factor for describing the full electrostatic potential is, however, $\Upsilon$ or equivalently $\Delta$.

Finally, in Fig. \ref{fig:parameters}(c) I plot a statistical parameter that quantifies the quality of the surface potential fit, the coefficient of determination or $\mathcal{R}$-squared value $\mathcal{R}^2$, with $\mathcal{R}^2=1$ indicating a perfect fit. From Fig. \ref{fig:parameters}, one can see that the fit is of lesser quality for larger $\kappa R_O$ and thicker particles, which is not surprising considering that an infinitely thin torus would be described perfectly by the mapping, and thick particles are a deviation from that. However, the quality is good enough to approximate the full spatial dependence of the electrostatic potential; see also the next section.

\section*{Appendix C: More comparisons of the electrostatic potential for an electrostatically screened charged torus}
\label{sec:morepot}

\renewcommand{\theequation}{C.\arabic{equation}}
\renewcommand{\thefigure}{C\arabic{figure}}
\setcounter{figure}{0}
\setcounter{equation}{0}

\begin{figure*}[h]
\centering
\includegraphics[width=\textwidth]{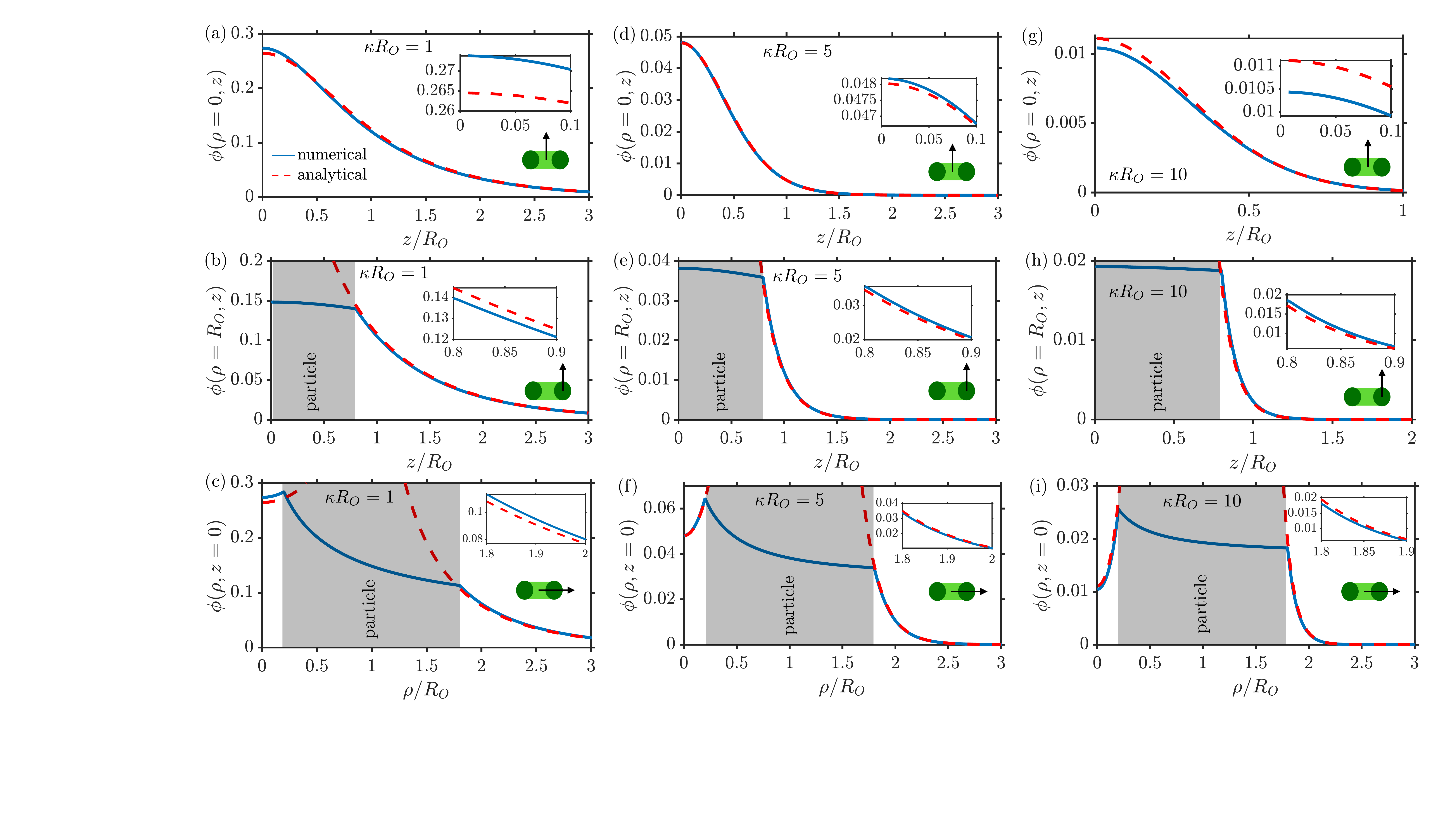}
\caption{Comparison of electrostatic potential of the semi-analytical ring-charge mapping with numerical finite-element calculations of the Poisson-Boltzmann equation for a toroidal particle at various values of $\kappa R_O$, along various cuts around the torus as shown in the insets, as well as an enlargement of the comparison between numerical and analytical results close to the particle surface. In all plots, $R_I/R_O=0.8$, which is a thicker torus than the one discussed in the main text (Fig. 1). Furthermore, $\epsilon_p/\epsilon_s=0.2$, $\ell_B/R_O=0.01$ and $Z=50$. The values of $\Delta$, or equivalently $\Upsilon$, and $R$ are determined from Fig. \ref{fig:fits}.}
\label{fig:thicktori}
\end{figure*}

In Fig. \ref{fig:toruspot} of the main text, I showed some comparisons of the numerically obtained electrostatic potential with the semianalytical approximated ring-charge mapping. Here, I will show that the approximation works even in a relatively ``extreme" cases of an even thicker torus $R_I/R_O=0.8$, and extremely thin double layers (up until $\kappa R_O=10$). The results are shown in Fig. \ref{fig:thicktori}. Regarding the charge parameters and shape parameters for these tori I find for $\kappa R_O=1$, $R=1.01058$ and $\Delta=1.1879$; for $\kappa R_O=5$, I find $R=1.0545$ and $\Delta=1.8078$; and finally, for $\kappa R_O=10$, I find $R=1.0392$ and $\Delta=2.2751$. See also Fig. \ref{fig:parameters}. 

\section*{Appendix D: Derivation of the torus anisotropy function with Yukawa multipole expansions}
\renewcommand{\theequation}{D.\arabic{equation}}
\renewcommand{\thefigure}{D\arabic{figure}}
\setcounter{figure}{0}
\setcounter{equation}{0}
Sometimes it is beneficial to assess the far-field behaviour of the electrostatic potential, which can be obtained with multipole expansion. Such an expansion is useful in deriving, for example, the anisotropy function of a particle; see Eq. \eqref{eq:anisofunc} in the main text. For the dimensionless electrostatic potential of the form 
\begin{equation}
\varphi({\bf r})=\ell_B\int d{\bf r}'\, q({\bf r}')G({\bf r},{\bf r}'),
\end{equation}
a general multipole expansion is given by
\begin{equation}
\varphi({\bf r})=\ell_B\sum_n\frac{1}{n!}{G}_{i_1,...,i_n}^{(n)}({\bf r}){T}_{i_1,...,i_n}^{(n)}=\ell_B\sum_n\frac{1}{n!}\frac{\partial^n}{\partial r_{i_1}'...\partial r_{i_n}'}G({\bf r},{\bf r}')\Big{|}_{{\bf r}'=0}\int d {\bf r}' \, r_{i_1}'...r_{i_n}'q({\bf r}').
\end{equation}
Here ${\bf G}^{(n)}({\bf r})$ is the $2^n$-polar basis functions expressed in tensor form and ${\bf T}^{n}$ is the $2^n$-pole moment tensor. Both tensors are symmetric in all indices, and I used the Einstein summation convention. Now consider the DH Green's function $G({\bf r},{\bf r}')=\exp(-\kappa|{\bf r}-{\bf r}'|)/|{\bf r}-{\bf r}'|$. In this case, I find up until $n=4$,
\begin{gather}
{G}^{(0)}({\bf r})=\frac{\exp(-\kappa r)}{r}, \\
{G}^{(1)}_i({\bf r})=\frac{\exp(-\kappa r)}{r^3}\left(1+\kappa r\right){r_i}, \\
{G}^{(2)}_{ij}({\bf r})=\frac{\exp(-\kappa r)}{r^5}\Big\{\left[3+3\kappa r+(\kappa r)^2 \right]r_ir_j-r^2(1+\kappa r)\delta_{ij}\Big\}, \\
{G}^{(3)}_{ijk}({\bf r})=\frac{\exp(-\kappa r)}{r^7}\Big\{\left[(\kappa r)^3+6(\kappa r)^2+15\kappa r+15\right]r_ir_jr_k-r^2\left[(\kappa r)^2+3\kappa r+3\right](r_i\delta_{jk}+r_j\delta_{ik}+r_k\delta_{ij})\Big\},
\end{gather}
\begin{align}
{G}_{ijkl}^{(4)}({\bf r})=&\frac{\exp(-\kappa r)}{r^9}\Big\{\left[105+105(\kappa r)+45(\kappa r)^2+10(\kappa r)^3+(\kappa r)^4\right]r_ir_jr_kr_l\nonumber \\
&-r^2\left[15+15\kappa r+6(\kappa r)^2+(\kappa r)^3\right](r_ir_j\delta_{kl}+r_kr_l\delta_{ij}+r_ir_k\delta_{jl}+r_jr_l\delta_{ik}+r_ir_l\delta_{jk}+r_jr_k\delta_{il})\nonumber \\
&+r^4\left[(\kappa r)^2+3\kappa r+3\right](\delta_{ij}\delta_{kl}+\delta_{ik}\delta_{jl}+\delta_{il}\delta_{jk})\Big\},
\end{align}
with ${\bf G}^{(0)}({\bf r})$ being the monopolar, ${\bf G}^{(1)}({\bf r})$ the dipolar, ${\bf G}^{(2)}({\bf r})$ the quadrupolar, ${\bf G}^{(3)}({\bf r})$ the octapolar, and ${\bf G}^{(4)}({\bf r})$ the hexadecapolar basis function tensors, respectively. Note that there are some differences with the unscreened case; for example, the quadrupolar tensor is not traceless, and therefore ${\bf T}^{(2)}$ cannot be chosen traceless.
For line charge distributions, the $2^n$-pole moment tensors simplify to
\begin{equation}
{T}_{i_1...i_n}^{(n)}=\int_a^b du \, \lambda(u)|\boldsymbol{\gamma}'(u)| \gamma_{i_1}(u)...\gamma_{i_n}(u).
\end{equation}
Specifically, for a uniformly charged ring, $\lambda=Q_r/(2\pi R)$, ${\boldsymbol{\gamma}}(u)=R(\cos u,\sin u,0)$, I find up until hexadecapolar order, 
\begin{gather}
{T}^{(0)}=Q_r, \\
{T}_{ij}^{(2)}=\frac{Q_r R^2}{2}\left(\delta_{ij}-\delta_{iz}\delta_{jz}\right), \\
{T}_{ijkl}^{(4)}=\frac{Q_r R^4}{8}[\delta_{ij}\delta_{kl}+\delta_{ik}\delta_{jl}+\delta_{il}\delta_{jk}+3\delta_{iz}\delta_{jz}\delta_{kz}\delta_{lz}-(\delta_{iz}\delta_{jz}\delta_{kl}+\delta_{kz}\delta_{lz}\delta_{ij}\nonumber \\+\delta_{iz}\delta_{kz}\delta_{jl}+\delta_{jz}\delta_{lz}\delta_{ik}+\delta_{iz}\delta_{lz}\delta_{jk}+\delta_{jz}\delta_{kz}\delta_{il})],
\end{gather}
and ${\bf T}^{(n)}=0$ for $n$ odd.
\begin{figure*}[t]
\centering
\includegraphics[width=0.8\textwidth]{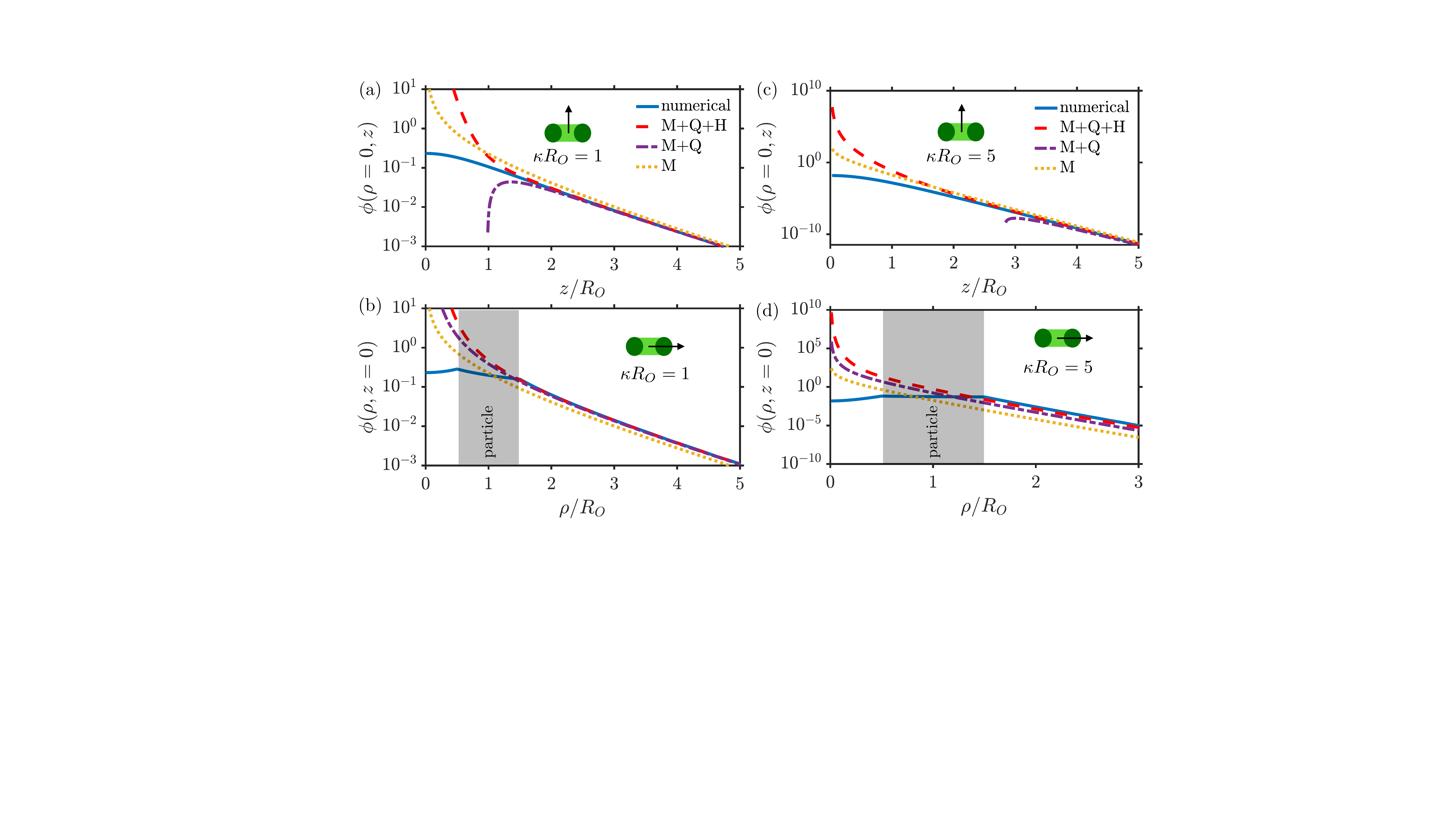}
\caption{Comparison of the electrostatic potential of a charged torus as obtained from finite-element calculations, with the Yukawa multipolar-expanded form Eq. \eqref{eq:multipole} on log-linear scale. The contributions of the various multipoles are highlighted, where monopolar (M), quadrupolar (Q), and hexadecapolar (H) effects are considered. The blue lines are the same as the one discussed in the main text Fig. 1. For the purple lines, the values are cut off because for sufficiently small $z$ at $\rho=0$ the potential becomes negative up until quadrupolar order.}
\label{fig:multipolecompare}
\end{figure*}

It is then straightforward to calculate the tensor contractions for the monopole and quadrupole term. I find
\begin{gather}
G^{(0)}({\bf r})T^{(0)}=\frac{Q_r\exp(-\kappa r)}{r}, \\
G_{ij}^{(2)}({\bf r})T_{ij}^{(2)}=\frac{Q_r R^2}{2}\frac{\exp(-\kappa r)}{r^5}\Big\{\left[3+3\kappa r+(\kappa r)^2 \right](r^2-z^2)-2r^2(1+\kappa r)\Big\}.
\end{gather}
 In order to compute the tensor contractions for the hexadecapolar term, it is useful to use the formula 
\begin{equation}
{G}_{i_1...i_n}^{(n)}({\bf r}){T}_{i_1...i_n}^{(n)}=\sum_{i=1}^n {{n}\choose{i}}G^{(n)}_{n-i,i}\,T^{(n)}_{n-i,i},
\end{equation}
which is valid because all $z$ components are zero for the ring. Furthermore, the notation $G^{(n)}_{a,b}$ means that $x$ occurs $a$ times and $y$ occurs $b$ times .
Specifically,
\begin{equation}
G^{(4)}_{ijkl}({\bf r})T^{(4)}_{ijkl}=G^{(4)}_{xxxx}({\bf r})T^{(4)}_{xxxx}+6G^{(4)}_{xxyy}({\bf r})T^{(4)}_{xxyy}+G^{(4)}_{yyyy}({\bf r})T^{(4)}_{yyyy},
\end{equation}
which results in
\begin{align}
G^{(4)}_{ijkl}({\bf r})T^{(4)}_{ijkl}=&\frac{Q_rR^4}{8}\frac{\exp(-\kappa r)}{r^9}\Big\{\left[105+105(\kappa r)+45(\kappa r)^2+10(\kappa r)^3+(\kappa r)^4\right]3(x^4+2x^2y^2+3y^4)\nonumber \\
&+24r^4\left[(\kappa r)^2+3\kappa r+3\right]-r^2\left[15+15\kappa r+6(\kappa r)^2+(\kappa r)^3\right]24(x^2+y^2)\Big\}.
\end{align}
Passing to spherical coordinates, I find up until hexadecapolar order
\begin{align}
\varphi(r,\theta)&=\frac{Z_t\ell_B\Upsilon\exp(-\kappa r)}{r}\Bigg(1+\frac{1}{4}\Big\{\left[3+3\kappa r+(\kappa r)^2 \right]\sin^2\theta-2(1+\kappa r)\Big\}\left(\frac{R}{r}\right)^2 \nonumber \\
&+\frac{1}{64}\Big\{\left[105+105(\kappa r)+45(\kappa r)^2+10(\kappa r)^3+(\kappa r)^4\right]\sin^4\theta+8\left[(\kappa r)^2+3\kappa r+3\right]\nonumber \\
&-\left[15+15\kappa r+6(\kappa r)^2+(\kappa r)^3\right]8\sin^2\theta\Big\}\left(\frac{R}{r}\right)^4+\mathcal{O}\left[\left(\frac{R}{r}\right)^6\right]\Bigg).\label{eq:multipole}
\end{align}
The accuracy of Eq. \eqref{eq:multipole} is tested against numerics in Fig. \ref{fig:multipolecompare}, showing that at least up until quadrupolar terms are needed to describe the far-field anisotropy sufficiently and hexadecapolar terms increase the accuracy even further at shorter distances from the particle. For $r\rightarrow\infty$,
\begin{align}
\varphi(r,\theta)\sim\frac{Z_t\ell_B\exp(-\kappa r)}{r}\underbrace{\Upsilon\left\{1+\frac{1}{4}\sin^2\theta\,(\kappa R)^2+\frac{1}{64}\sin^4\theta\,(\kappa R)^4+\mathcal{O}\left[(\kappa R)^6\right]\right\}}_{:=\mathcal{A}(\theta)},
\label{eq:anisofunc2}
\end{align}
\begin{figure*}[t]
\centering
\includegraphics[width=0.4\textwidth]{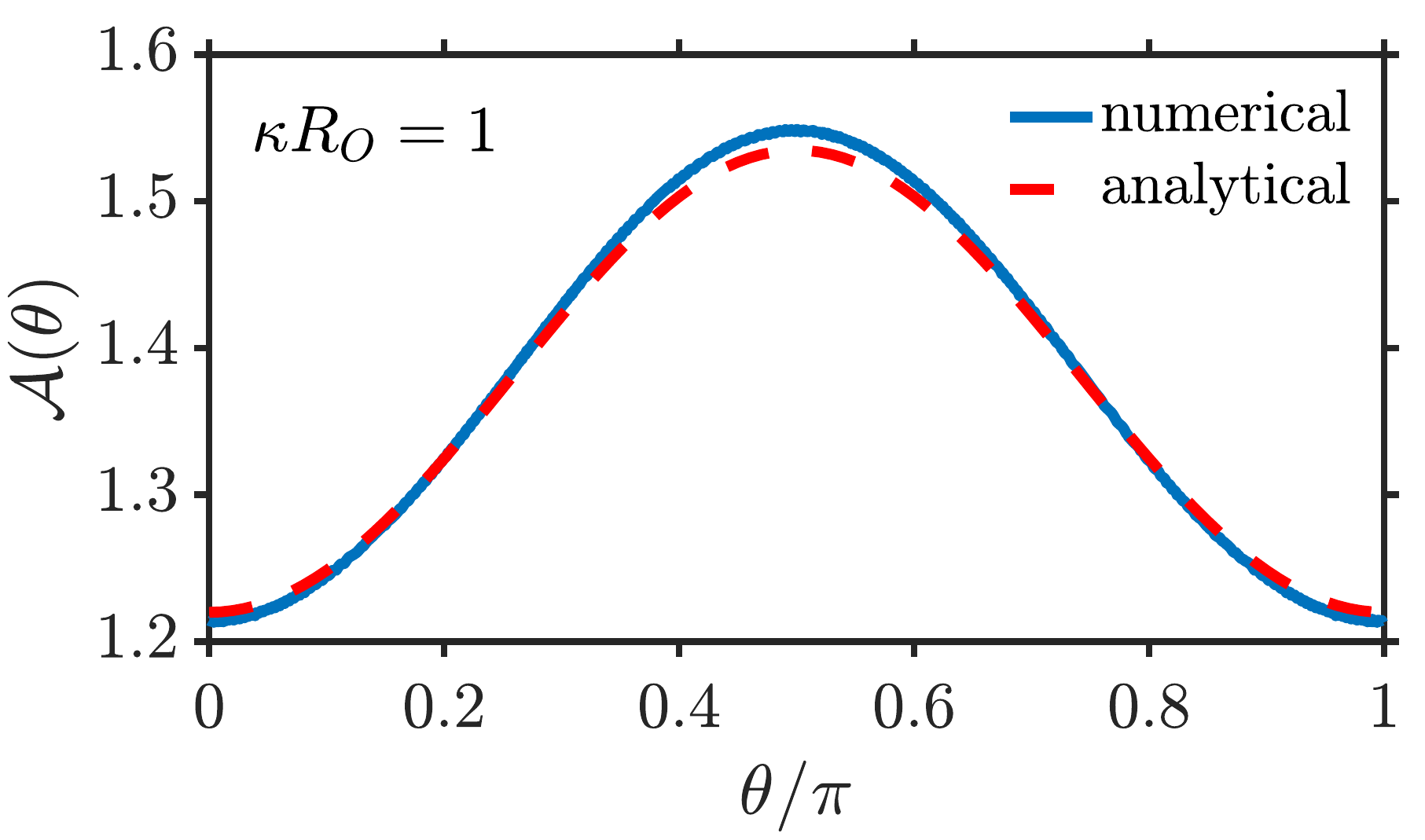}
\caption{Comparison of the numerically obtained anisotropy function with the multipolar expanded anisotropy function obtained from Eq. \eqref{eq:anisofunc}. The anisotropy function was obtained for the same parameters as the potential in main text Figs. 1(a, c-e).}
\label{fig:anisofunc}
\end{figure*}
where I defined the anisotropy function $\mathcal{A}(\theta)$. I compare $\mathcal{A}(\theta)$ in Fig. \ref{fig:anisofunc} for $\kappa R_O=1$ for the torus discussed in the main text. Unfortunately, it is difficult to make comparisons for shorter screening lengths. To appreciate why, consider the example that the anisotropy function from Eq. \eqref{eq:anisofunc} is not sensitive at $\kappa R_O=5$ for inclusion of the hexadecapolar term; therefore, I can suffice with only the quadrupolar term. In order to numerically compare Eq. \eqref{eq:anisofunc2} with the asymptotic expression for $r\rightarrow\infty$, one must have that $3\kappa R^2/r\ll 1$ (highest order in $1/r$ quadrupolar term after the constant term). Already at $3\kappa R=1$ the potentials are, however, $\mathcal{O}(10^{-8})$, so one needs to resolve numerically $\varphi({\bf r})\ll 10^{-8}$ which is difficult to do considering that the potential close to the surface is $\mathcal{O}(10^{-2})$ for the considered set of parameters; see Fig. \ref{fig:multipolecompare}.

\end{document}